\documentclass[longauth, letter]{aa}

\AtBeginDocument{}

\usepackage{graphicx}
\usepackage{txfonts}
\usepackage{lipsum}
\usepackage{subcaption}
\usepackage{lscape}
\usepackage{placeins}
\usepackage{orcidlink}
\usepackage{multirow}

\newcommand{\chg}[1]{{#1}}

\begin{document}

   \title{Probing the geometry and outflows of the accreting millisecond pulsar SAX J1808.4–3658 with multi-band polarimetry}

   \author{M. C. Baglio
          \inst{1}\orcidlink{0000-0003-1285-4057}
          \and
        G. Illiano\inst{2,3,1}\orcidlink{0000-0003-4795-7072}
        \and
          A. Papitto\inst{4}\orcidlink{0000-0001-6289-7413}    
          \and
          D. M. Russell\inst{5}\orcidlink{0000-0002-3500-631X}
          \and
          A. Di Marco\inst{6}
%          \and
%          A. Veledina\inst{6}
          \and
          K. Alabarta\inst{1}
          \and
          C. Ballocco\inst{4, 7}
          \and
          S. Campana \inst{1}
          \and
          S. Covino\inst{1,8}
          \and
          P. D'Avanzo \inst{1}
          \and
          R. P. Fender\inst{9, 10}
          \and
          K. Gasealahwe\inst{10,11}
          \and
          F. Lewis\inst{12,13}
          \and
          M. M. Messa\inst{1}
          \and
          S. Motta\inst{1}
          \and
          P. Saikia\inst{14}
          }          

         %Sasha, Sara, Rob ?
   \institute{INAF, Osservatorio Astronomico di Brera, Via E. Bianchi 46, I-23807 Merate (LC), Italy\\
              \email{maria.baglio@inaf.it}
              \and
              Institute of Space Sciences (ICE, CSIC), Campus UAB, Carrer de Can Magrans s/n, E-08193 Barcelona, Spain
              \and
              Institut d’Estudis Espacials de Catalunya (IEEC), E-08860 Castelldefels (Barcelona), Spain  
              \and
            INAF -- Osservatorio Astronomico di Roma, via di Frascati 33, I-00078 Monteporzio Catone (Roma), Italy
              \and
              Center for Astrophysics and Space Science, NYUAD Research Institute, New York University Abu Dhabi, PO Box 129188, Abu Dhabi, UAE
            \and
              INAF Istituto di Astrofisica e Planetologia Spaziali, Via del Fosso del Cavaliere 100, I-00133 Roma, Italy
              \and
              Dipartimento di Fisica, Sapienza Università di Roma, Piazzale Aldo Moro 5, I-00185 Rome, Italy
              \and
              Como Lake centre for AstroPhysics (CLAP), DiSAT, Universit\`a dell’Insubria, via Valleggio
              11, 22100 Como, Italy
              \and
              Astrophysics, Department of Physics, University of Oxford, Oxford, UK
              \and
              Department of Astronomy, University of Cape Town, Private Bag X3, Rondebosch 7701, South Africa 
              \and
              South African Astronomical Observatory, PO Box 9, Observatory, Cape
Town 7935, South Africa
            \and
%            Nordita, Stockholm University and KTH Royal Institute of Technology, Hannes Alfv\'ens v\"ag 12, SE-10691 Stockholm, Sweden
%              \and
            Faulkes Telescope Project, Cardiff, Wales, UK
            \and
            The Schools’ Observatory, Astrophysics Research Institute, Liverpool John Moores University, 146 Brownlow Hill, Liverpool L3
5RF, UK
\and
            Department of Astronomy, Yale University, PO Box 208101, New Haven, CT 06520-8101, USA
              }

   \date{}
\authorrunning{Baglio et al.}
\titlerunning{Optical Polarimetry of SAX J1808.4-3658 during the 2025 outburst}

  \abstract{We present VLT/FORS2 $R$- and $I$-band imaging polarimetry of the accreting millisecond X-ray pulsar SAX~J1808.4$-$3658 during its 2025 outburst, coordinated with the first X-ray polarimetric observations of this source with \textit{IXPE}. We detect significant optical linear polarisation at every epoch (polarisation degree $P\simeq0.3$--1.5\%, angle $\Phi\simeq8$--48$^\circ$), consistent with the only prior optical polarimetric campaign of this source, obtained during a previous outburst. The average optical polarisation angle is offset by $\sim16$--24$^\circ$ from the \textit{IXPE} angle of the persistent X-ray emission, which pulse-profile modelling shows to trace the spin axis. A fast-timing sequence reveals the optical polarisation angle swinging by up to $\sim40^\circ$ within minutes, disfavouring scattering in the disc. A colour-magnitude diagram and the dependence of the optical polarisation on the red colour excess over the disc emission, which we attribute to the jet, instead point to a jet origin during the main outburst, consistent with contemporaneous radio detections. The offset, with the same mean optical angle as in the previous outburst, can arise from emission near a wide jet base, where the Doppler-boosted side of the cone dominates. At the onset of the reflaring stage the optical polarisation degree is highest and its angle coincides with the \textit{IXPE} one, while the disc likely dominates the optical flux. We discuss a polarisation flare of the fading jet, scattering in an
outflow, and synchrotron emission at the disc--magnetosphere boundary
as possible origins, tentatively favouring the last.}
  

   \maketitle

%%%%%%%%%%%%%%%%%%%%%%%%%%%%%%%%%%%%%%%%%%%%%%%%%%%%%%%%%%%%%%
\section{Introduction}
SAX~J1808.4$-$3658 was the first accreting millisecond X-ray pulsar
(AMXP; see \citealt{Patruno_Watts_2021ASSL,DiSalvo_Sanna_2022ASSL,baglio2026} for most recent reviews) discovered \citep{Wijnands1998} and is the most frequently
recurring of its class, with eleven outbursts since 1996
\citep[e.g.][]{Goodwin2020, Illiano2023,Papitto2026}. Its 2.01\,hr orbit, 401\,Hz spin, and
persistent hard state make it a favourable target for probing the
accretion flow, its outflows, and the magnetosphere with
multi-wavelength polarimetry.

A near-infrared/optical excess above the reprocessed-disc spectrum has
long been attributed to optically thin synchrotron emission from a
compact jet \citep{Wang2001,Greenhill2006,Baglio2015,Baglio2020}.
During the 2019 outburst, \citet{Baglio2020} detected variable optical linear polarisation, with degree $P=0.2$--$2.2\%$ and angle $\Phi=0$--50$^\circ$ (measured from north through east), which they attributed to a jet with a strongly tangled magnetic field, based mainly on a positive correlation between $P$ and the jet contribution to the optical flux.

During the 2025 outburst, \textit{IXPE} detected polarisation in the persistent 2--8\,keV emission (\chg{$P=2.9\pm0.5\%$,
$\mathrm{PA}=46\pm5$\degr}; hereafter we use PA for the X-ray and $\Phi$ for the optical polarisation angle) and, tentatively, during a type-I burst, with a higher degree ($15.5\pm6.1\%$) and an angle ($\simeq-44^\circ$) rotated by $\sim90^\circ$ from the persistent one \citep[][hereafter P26]{Papitto2026}. P26
also derive the sky-projected spin-axis angle from pulse-profile
modelling, \chg{$\chi_p=41\pm5$\degr}, aligned with the persistent PA as
expected for Comptonisation in the accretion column
\citep{Bobrikova2023}. Here we present VLT/FORS2
\textit{R}- and \textit{I}-band polarimetry spanning the outburst rise, the early decay, and the onset of the reflaring stage.

%%%%%%%%%%%%%%%%%%%%%%%%%%%%%%%%%%%%%%%%%%%%%%%%%%%%%%%%%%%%%%
\section{Observations and data analysis}
\label{sec:obs}
%%%%%%%%%%%%%%%%%%%%%%%%%%%%%%%%%%%%%%%%%%%%%%%%%%%%%%%%%%%%%%

\label{sec:polobs}
We observed SAX~J1808.4$-$3658 with VLT/FORS2 in imaging polarimetry
mode (Wollaston prism and half-wave plate at $0$, $22.5$, $45$,
$67.5$\degr; Programme 115.28AY.001, PI: Baglio). We obtained three
single-epoch \textit{R}+\textit{I} observations (2025 Aug 11/12,
Aug 20/21, Sep 4/5; orbital phases \chg{$\psi_{\rm orb}$}$\simeq0.15$, $0.69$, $0.81$
with the P26 ephemeris, their Table~1: $T_{\rm asc}={\rm MJD}\,60907.052902$,
$P_{\rm orb}=7249.154$\,s) and an 11-sub-epoch \textit{R}-band sequence
over $\sim\!62$\,min on Aug 12/13 (\chg{$\psi_{\rm orb}$}$\simeq0.83$--$1.28$, wrapping
through the ascending node); the log is given in Table~\ref{tab:log}.
On each single-epoch night, the $R$ and $I$ sequences are consecutive,
not simultaneous. $P$ and $\Phi$ were derived with a Bayesian fit of
the $S$-parameter \citep{Baglio2020}, after correcting for
instrumental and interstellar polarisation with a set of field
reference stars, pooled across the three single-epoch nights. Details
of the reduction, calibration, and reference-star baseline are given
in Appendix~\ref{app:polmethod}.

\label{sec:monitoring}
We also use $V$-, $i'$-, and $R$-band monitoring from the Las Cumbres
Observatory (LCO) network \citep{Lewis2008}, reduced and analysed with the \textsc{XB-NEWS} pipeline \citep[][Appendix~\ref{app:lco}]{Russell2019,Saikia2026}.
These observations provide the colour-magnitude diagram (see Appendix~\ref{app:cmd}) and the absolute flux
calibration of the FORS2 target fluxes (Appendix~\ref{app:polflux}). Figure~\ref{fig:lc} shows the $i'$-band light
curve with the polarimetric epochs highlighted by black solid lines.

%%%%%%%%%%%%%%%%%%%%%%%%%%%%%%%%%%%%%%%%%%%%%%%%%%%%%%%%%%%%%%
\section{Results}
\label{sec:results}
%%%%%%%%%%%%%%%%%%%%%%%%%%%%%%%%%%%%%%%%%%%%%%%%%%%%%%%%%%%%%%

\begin{table}
\caption{Polarisation results.}
\label{tab:results}
\centering
\begin{tabular}{lccc}
\hline\hline
Epoch & Band & $P$ (\%) & $\Phi$ (\degr)\\
\hline
\multirow{2}{*}{Aug 11/12} & $R$ & $0.65\pm0.07$ & $30.8\pm3.0$\\
                           & $I$ & $0.70\pm0.06$ & $24.4\pm2.2$\\
\noalign{\smallskip}
Aug 12/13 (01--11)\chg{\tablefootmark{a}}        & $R$ & $0.34$--$0.74$ & $8.6$--$47.8$\\
\noalign{\smallskip}
\multirow{2}{*}{Aug 20/21} & $R$ & $0.84\pm0.07$ & $36.9\pm2.4$\\
                           & $I$ & $0.70\pm0.06$ & $20.1\pm2.5$\\
\noalign{\smallskip}
\multirow{2}{*}{Sep 4/5}   & $R$ & $1.54^{+0.18}_{-0.19}$ & $47.2\pm3.4$\\
                           & $I$ & $<0.98$ & --\\
\hline
\textit{IXPE} T1\tablefootmark{b} & 2--8\,keV & \chg{$2.9\pm0.5$} & \chg{$46\pm5$}\\
\hline
\end{tabular}
\tablefoot{Upper limits are $3\sigma$ (Appendix~\ref{app:Smethod}).
\tablefoottext{a}{\chg{Range of the median values of the 11 sub-epochs (Fig.~\ref{fig:phase}), 
excluding their uncertainties; individual $1\sigma$ uncertainties are 
$\simeq 0.08$--$0.13\%$ in $P$ and $\simeq 3$--$8^\circ$ in $\Phi$.}}
\tablefoottext{b}{Persistent emission during the outburst decay (interval
T1 of P26); the angle is the X-ray PA.}
}
\end{table}

\begin{figure}
\centering
\includegraphics[width=\hsize]{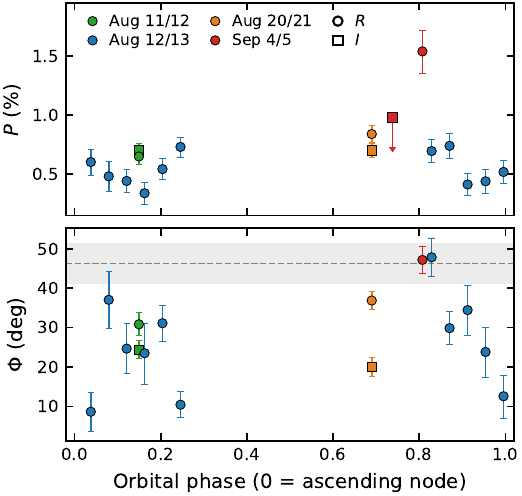}
\caption{Optical polarisation $P$ (top) and angle $\Phi$ (bottom) versus orbital phase, colour-coded by epoch ($R$: circles; $I$: squares; arrow: $3\sigma$ upper limit). Grey band and dashed line: \textit{IXPE} T1 PA ($1\sigma$ range and central value).}
\label{fig:phase}
\end{figure}

Polarisation is detected at every epoch (Table~\ref{tab:results}; values
corrected for instrumental and interstellar polarisation, Appendices~\ref{app:instpol} and
\ref{app:ism}), with $P=0.3$--$1.5\%$ and $\Phi=8$--$48$\degr,
consistent with 2019 \citep{Baglio2020}; the only non-detection is the
Sep~4/5 $I$ band ($3\sigma$ upper limit $0.98\%$). Both $P$ and $\Phi$
vary substantially between epochs, with no clear phase-locked
pattern (Fig.~\ref{fig:phase}). An $F$-test comparing a single sinusoid
at the orbital period with a constant, fitted to $P$ and $\Phi$ of the
14 $R$-band epochs, gives only a marginal improvement ($F=3.98$,
$p=0.050$ for $P$; $F=4.23$, $p=0.044$ for $\Phi$; 2 and 11 degrees of
freedom). We do not consider this convincing, given the scatter and
that \chg{$\psi_{\rm orb}$}$\simeq0.25$--$0.6$ is unsampled.

The $R$ and $I$ angles agree at Aug 11/12 but differ by $\sim\!17$\degr\
($\sim\!5\sigma$) at Aug 20/21, a discrepancy persisting after
pooling the reference stars (Appendix~\ref{app:pool}). Over the two
main-outburst epochs, the weighted mean angles are
$\overline{\Phi}_R=34.6\pm1.9$\degr\ and
$\overline{\Phi}_I=22.5\pm1.7$\degr\ ($4.9\sigma$ apart), offset from
the \textit{IXPE} PA by \chg{$11.4\pm5.3$\degr}\ ($2.1\sigma$) and
\chg{$23.5\pm5.3$\degr}\ ($4.4\sigma$), respectively: neither optical band
coincides with the X-ray angle, and their offsets differ. The average
over all 14 $R$-band epochs, $\overline{\Phi}_R=30.3\pm1.1$\degr, is
offset by \chg{$15.7\pm5.1$\degr}\ ($\sim\!3\sigma$). Only two individual
measurements coincide with the \textit{IXPE} angle: epoch~01 of
Aug~12/13 ($47.8\pm4.9$\degr) and Sep~4/5 in $R$ ($47.2\pm3.4$\degr),
22~days apart but at similar orbital phase (\chg{$\psi_{\rm orb}$}$\simeq0.83$, $0.81$).
%%%%%%%%%%%%%%%%%%%%%%%%%%%%%%%%%%%%%%%%%%%%%%%%%%%%%%%%%%%%%
\section{Discussion and conclusions}
\label{sec:discussion}
%%%%%%%%%%%%%%%%%%%%%%%%%%%%%%%%%%%%%%%%%%%%%%%%%%%%%%%%%%%%%%

To identify the component dominating the optical polarised flux, we first consider scattering in the disc atmosphere. 
%Several arguments disfavour it as the dominant source of the optical polarisation. 
This is a stable, axis-aligned mechanism: it would keep $\Phi$ fixed regardless of the varying contribution of other components, with $P$ set by dilution, and potentially add an orbital modulation as in the black hole X-ray binary GRO J1655$-$40 and the neutron star X-ray binary Her X$-$1 \citep{Gliozzi98,Schultz04}, which we do not find here (Sect.~\ref{sec:results}). Instead, during the fast-timing sequence of Aug 12/13, $\Phi$ swings between 8.6 and 47.8$^\circ$ within minutes: scattered light follows changes in the illumination almost instantaneously ($r_{\rm out}/c\sim1$\,s), but a stationary, axisymmetric disc still produces a fixed angle set by its axis. A disc origin for the observed polarisation would also imply a spin-orbit misalignment, given the $\sim3\sigma$ offset of $\overline{\Phi}_R$ from the \textit{IXPE} angle. %Its polarisation level is discussed below, after introducing the disc contribution to the flux.

\begin{figure}
    \centering
    \includegraphics[width=1\linewidth]{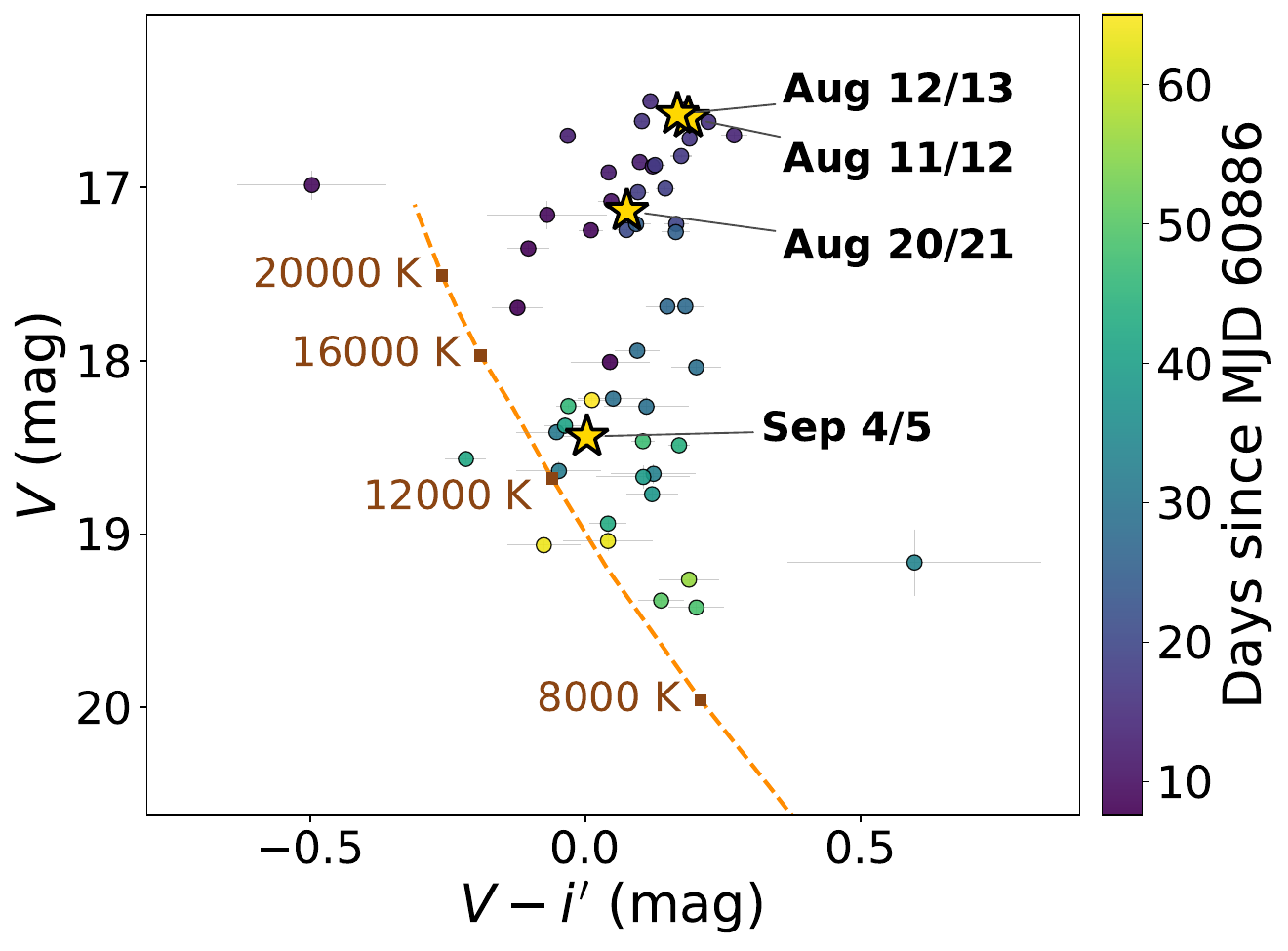}
    \caption{$V$ vs.\ $V-i'$ diagram of the 2025 outburst. Circles: LCO $V$/$i'$ pairs (one per night, taken within 1\,d), colour-coded by days since MJD 60886 (outburst onset). Gold stars: the pairs closest in time to the four FORS2 epochs. Dashed line: blackbody disc track of \citet{Baglio2020}, with selected temperatures.}
    \label{fig:cmd}
\end{figure}

The disc nevertheless contributes much of the optical flux. In a $V$ vs. $(V-i')$ colour-magnitude diagram (CMD) built from nightly-averaged LCO photometry and compared with the single-temperature blackbody disc track of \citet{Baglio2020} (Fig.~\ref{fig:cmd}; Appendix~\ref{app:cmd}), the three main-outburst epochs lie well redward of the track, whereas Sep 4/5, at the reflare onset, falls close to it, with a small residual red excess as in 2019. Dominating the flux, however, does not imply dominating the polarised flux: at Sep 4/5, $F_{\rm pol,R}=3.72\,\mu$Jy is only 1.5\% of the total, which a component with $P\sim10$--20\% and $\sim10\%$ of the flux would supply. The polarised-flux spectrum cannot discriminate between components, since non-simultaneous $R$ and $I$ exposures yield both extremely blue (Sep 4/5) and extremely red (2019 epoch 6) indices in this rapidly flickering source (Appendix~\ref{app:polSED}).

\begin{figure}
    \centering
    \includegraphics[width=1\linewidth]{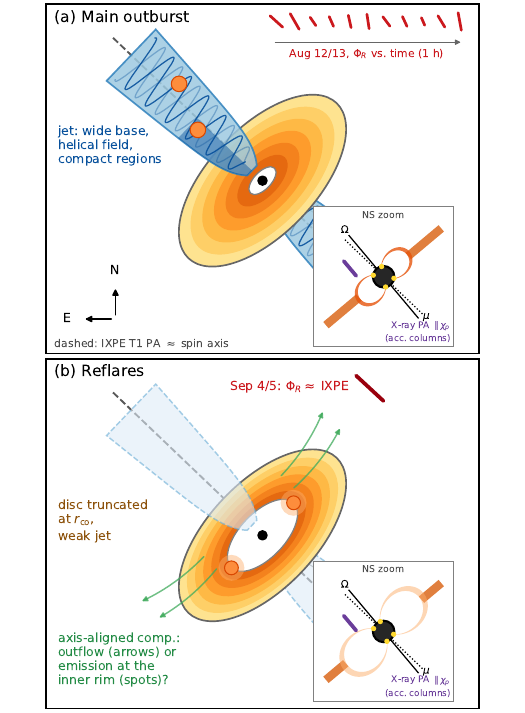}
    \caption{Sketch of the proposed geometry (not to scale; North up, East left). Red bars: measured $R$-band polarisation vectors, length
$\propto P$. Dashed line: \textit{IXPE} T1 PA ($\simeq$ spin axis).
(a) Main outburst: jet with a wide base (darker shading: side rotating towards the observer, Doppler-boosted) and compact emitting regions in a helical field. (b) Reflares:
disc truncated near $r_{\rm co}$, weak jet, and the candidate components
aligned with \textit{IXPE}: an equatorial outflow (green arrows) or
emission where the inner disc interacts with the magnetosphere (orange spots).
Insets: accretion columns, with X-ray PA $\parallel$ spin axis
($\chi_p$; P26).}
    \label{Fig:combined}
\end{figure}

\begin{figure}
    \centering
    \includegraphics[width=1\linewidth]{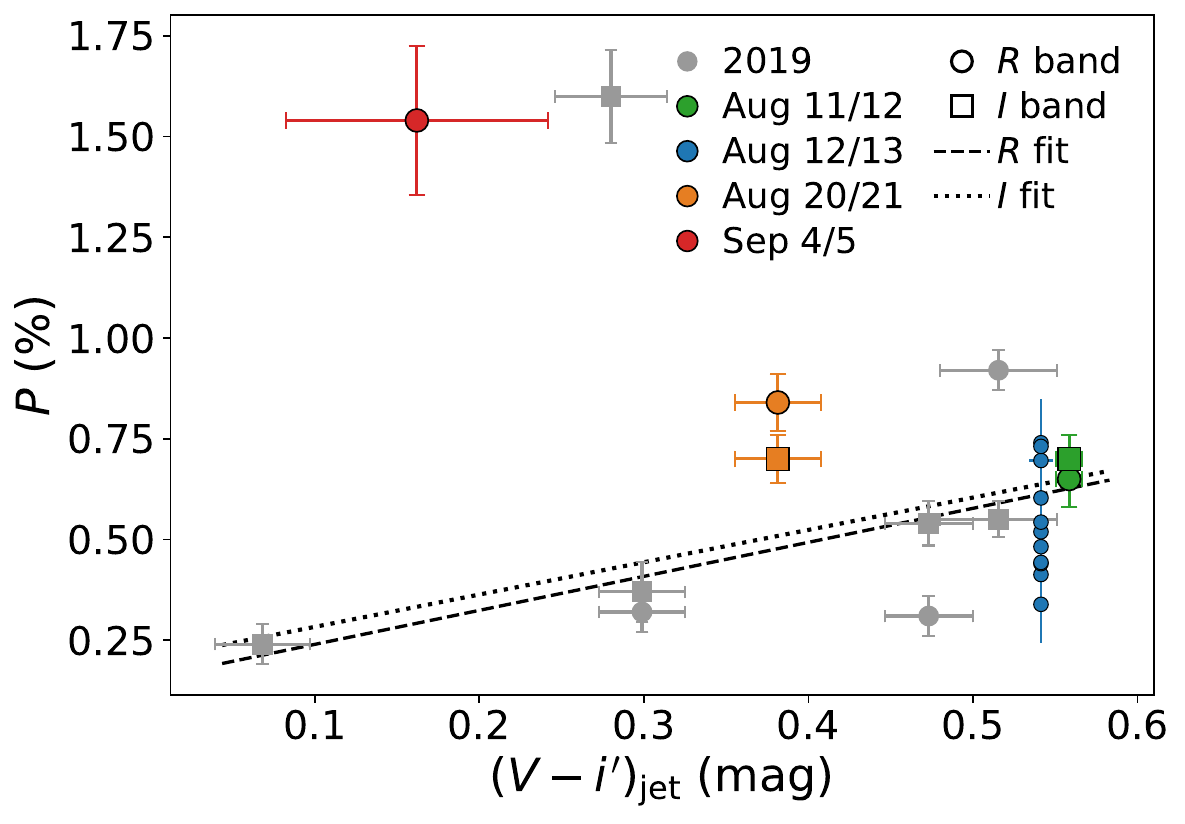}
    \caption{$P$ vs.\ $(V\!-\!i')_{\rm jet}$ for the 2019 (grey; \citealt{Baglio2020}) and 2025 outbursts (colours as in Fig.~\ref{fig:phase}; small circles: Aug~12/13 sub-epochs). Lines: weighted linear fits to the $R$ (dashed; slope $0.84\pm0.22$) and $I$ points (dotted; $0.80\pm0.13$), excluding Sep~4/5 ($R$) and 2019 epoch~6 ($I$).}
    \label{fig:V-i}
\end{figure}

We propose that, during the main outburst, the polarisation is instead dominated by the jet (Fig. \ref{Fig:combined}). The flickering of the polarisation is naturally produced by a jet with a tangled field \citep{Shahbaz2019}, as found in 2019 \citep{Baglio2020}. The red excess over the disc track also points to a significant jet contribution, consistent with the MeerKAT detections of a compact jet close
in time to our polarimetric epochs (Gasealahwe et al., in prep.). Subtracting the disc contribution from the observed colour, we reconstructed the $P$ vs. $(V-i')_{\rm jet}$ relation (Fig.~\ref{fig:V-i}), where $(V-i')_{\rm jet}$ is the colour excess relative to the disc track: most 2025 points follow the positive trend found in 2019, independently supporting a jet origin. Given the flickering, only the trend traced by the bulk of the points is informative.
The same relation constrains the polarisation of the disc. Were disc scattering the more polarised component, $P$ would increase towards the disc track; instead, it decreases. The 2019 epochs closest to the track have $P\simeq0.2$--0.3\%, comparable to the $\sim0.15\%$ measured in the disc-dominated soft state of the black hole X-ray binary MAXI J1820+070 \citep{Kosenkov2020}.

The polarisation angles also constrain the jet magnetic field. A purely disordered field would scatter $\Phi$ over 180$^\circ$, whereas our angles cluster around the same mean ($\sim30^\circ$) found in 2019, requiring an ordered component that persists between the two outbursts. For an axisymmetric jet with an ordered field, optically thin synchrotron emission, as expected in the optical, is polarised parallel (toroidal field) or perpendicular (poloidal field) to the projected jet axis, so the $\sim3\sigma$ offset of $\overline{\Phi}_R$ from the \textit{IXPE} angle requires the net polarised emission to depart from this symmetry. This does not necessarily imply a tilted jet. The optical emission presumably arises near the jet base, where the outflow is still conical or parabolic before recollimating (\citealt{Russell2014}; see also \citealt{Lu2023} for M87), so that the local field is inclined to the jet axis. The rotation the base inherits from the disc ($v_\phi\simeq0.3c$ at $\sim20$--30\,km) Doppler-boosts the side of the cone rotating towards the observer, tilting the net angle towards the field direction on that side, by up to the half-opening angle, while the collimated jet downstream remains aligned with the spin axis. Since the sense of the tilt is set by the disc rotation, such an offset naturally persists across epochs and between outbursts, as observed. A helical field sampled by a few compact emitting regions near the base, each covering only part of the helix \citep{Lyutikov2005}, would be consistent with the observed flickering around this mean (Fig.~\ref{Fig:combined}). Alternatively, the jet may trace a disc axis misaligned from the spin, as suggested for Cir X-1 \citep{Rankin2024}, although the angle variations there may equally reflect orbital changes in the mix of spectral components. Such a misalignment may not survive the long accretion history that aligned the magnetic and spin axes \citep{Lamb2009}, and a jet precessing around the spin axis, e.g. through Lense--Thirring precession of a geometrically thick inner flow, would make the angle drift around the \textit{IXPE} one rather than hold a fixed offset. A jet launched from a magnetically warped inner disc is discussed in Appendix~\ref{app:warp}.

At Sep~4/5 the picture changes. $P_R$ reaches its highest value, while in $I$ we obtain only a $3\sigma$ upper limit ($P_I<0.98\%$); $\Phi_R$ aligns with the \textit{IXPE} angle ($47.2\pm3.4^\circ$ vs. \chg{$46\pm5^\circ$}, otherwise seen only at epoch 01 of Aug 12/13), and this is the only 2025 epoch departing from the $P$--$(V-i')_{\rm jet}$ trend. Although it lies closest to the disc track, its polarisation is several times that of the 2019 epochs near the track, disfavouring the disc. The source had entered the reflaring stage, in which the disc may be truncated near the co-rotation radius, $r_{\rm co}=(GMP_{\rm spin}^2/4\pi^2)^{1/3}\simeq31$\,km for $M=1.4\,M_\odot$, in a propeller or trapped-disc configuration \citep{Patruno2016}, and the jet weakens (in 2019 the reflare radio flux stayed $\lesssim0.16$\,mJy\,beam$^{-1}$, against a main-outburst peak of 0.59\,mJy\,beam$^{-1}$; \citealt{Gasealahwe2023}). We consider three possible origins. (i) A polarisation flare of the fading jet, as in the 2019 reflares, where epoch 6 reached $\sim1.6\%$ \citep{Baglio2020}; this requires the alignment with \textit{IXPE} to be a coincidence, and a red flare should be at least as strong in $I$, unless it developed after the $I$-band exposure, which preceded the $R$-band one. (ii) Scattering in an optically thin wind or propeller-driven outflow, which would yield an angle tied to the orbital axis. A propeller transition is reported from \textit{NuSTAR} data of Sep 14 \citep{Bhattacharya2026}; with no earlier coverage of the reflares, a propeller-driven outflow may already have been present on Sep 4/5. Optical polarisation of $\sim0.3$--1\% has been attributed to outflows in the black hole binaries V404 Cyg \citep{Kosenkov2017}, MAXI J1820+070 \citep{Kosenkov2020}, and Swift J1727.8$-$1613, where its offset from the jet axis was taken as evidence of a spin-orbit misalignment \citep{Nitindala2026}; here, the alignment would instead require aligned orbital and spin axes. (iii) Synchrotron emission from particles accelerated where the inner
disc interacts with the magnetosphere, a process proposed for the
optical pulsations of this source \citep{Ballocco2026arXiv} and
analogous to the pulsar wind--disc boundary in the transitional
millisecond pulsar PSR~J1023+0038 \citep{Baglio2025}, with an angle
fixed relative to the projected spin axis. The inner disc recedes
during the reflares, and in 2019 the optical pulsed flux was
comparable to our polarised flux (Appendix~\ref{app:magneto}). We
favour (iii): it ties the angle directly to the spin axis, $P_R$
exceeds the level attributed to outflows, and the fluxes are
compatible, although whether this emission is optically thin remains
to be shown.

Several observations can discriminate between these scenarios.
Strictly simultaneous multi-band polarimetry would remove the
ambiguity of the polarised-flux spectrum, and polarimetry on
timescales of seconds would distinguish the stochastic flickering of
compact jet regions from the periodic swing expected from a precessing
warp (Appendix~\ref{app:warp}). For scenario (iii), simultaneous
optical polarimetry and fast photometry would test whether the
polarised and pulsed fluxes share a common origin, and instruments
able to measure the polarisation of the optical pulsations, such as
eSiFAP\footnote{\url{https://www.tng.iac.es/news/2025/02/24/first-on-sky-tests-of-esifap-fast-photometer/}},
would probe it directly.
%%%%%%%%%%%%%%%%%%%%%%%%%%%%%%%%%%%%%%%%%%%%%%%%%%%%%%%%%%%%%%
\begin{acknowledgements}
      Based on observations collected at the European Southern Observatory under ESO programme 115.28AY.001.
      %This research was supported by the International Space Science Institute (ISSI) in Bern, through ISSI International Team Project 'Maximizing the Potential of X-ray Polarimetric Data to Understand Accreting BHs' (ISSI Team project $\#25-649$).
MCB acknowledges support from the INAF Fundamental Research Grant XBOOM and from the L'Oréal-UNESCO For Women in Science Programme – Italy, through the 2026 Fellowship Award. GI is supported by a Juan de la Cierva fellowship (JDC2024-053550-I). DMR is supported by Tamkeen under the NYUAD Research Institute grant CG014. This work was supported by INAF Research Grant “PULSE-X” and by Fondazione Cariplo/CDP through Grant 2023-2560 (PI: Papitto). We thank Daniel Bramich for developing the XB-NEWS pipeline, on whose results part of this work is based. We thank A. Veledina for fruitful discussion that helped with the interpretation.
\end{acknowledgements}

%%%%%%%%%%%%%%%%%%%%%%%%%%%%%%%%%%%%%%%%%%%%%%%%%%%%%%%%%%%%%%
\bibliography{bibliography}{}
\bibliographystyle{aa}

\begin{appendix}

%%%%%%%%%%%%%%%%%%%%%%%%%%%%%%%%%%%%%%%%%%%%%%%%%%%%%%%%%%%%%%
\section{VLT/FORS2 polarimetry}
\label{app:polmethod}

\subsection{Observations and data reduction}
\label{app:polred}

All observations were carried out with the FORS2 filters
\texttt{R\_SPECIAL+76} ($R$; central wavelength 655\,nm, FWHM 165\,nm)
and \texttt{I\_BESS+77} ($I$; 768\,nm, FWHM 138\,nm). A Wollaston
prism splits the incoming light into two orthogonally polarised beams
(ordinary, $o$, and extraordinary, $e$), and a Wollaston mask prevents
the two beams from overlapping on the detector. A rotating half-wave
plate (HWP) was used to acquire images at four angles,
$\phi_i=22.5\degr(i-1)$, $i=1,\dots,4$. Exposure times were
35\,s in $R$ and 70\,s in $I$ per HWP angle for the three
single-epoch nights; the Aug~12/13 sequence consists of 11 consecutive
$R$-band HWP cycles of $4\times30$\,s, each cycle defining one
sub-epoch of $\sim\!5.6$\,min. On each single-epoch night the $R$- and
$I$-band sequences were taken one after the other, so that the two bands are not exactly simultaneous. Sky
conditions were good, with seeing below 1'' all the time. A log of the observations is presented in Tab. \ref{tab:log}.

\begin{table*}
\caption{VLT/FORS2 polarimetric observing log. $t_{\rm exp}$ is the
exposure time per HWP angle.}
\label{tab:log}
\centering
\begin{tabular}{lcccc}
\hline\hline
Night & MJD & Band & $N_{\rm epoch}$ & $t_{\rm exp}$ (s)\\
\hline
2025 Aug 11/12 & 60899.0806--60899.0952 & $R$, $I$ & 1  & 35, 70\\
2025 Aug 12/13 & 60899.9856--60900.0286 & $R$      & 11 & 30\\
2025 Aug 20/21 & 60908.1044--60908.1171 & $R$, $I$ & 1  & 35, 70\\
2025 Sep 4/5   & 60923.1319--60923.1464 & $R$, $I$ & 1  & 35, 70\\
\hline
\end{tabular}
\end{table*}

All images were bias-subtracted and flat-fielded, and aperture
photometry of the target and of the reference stars was performed on
both beams with \textsc{daophot} \citep{Stetson1987}, with an aperture
of $1.5\times FWHM$.

\subsection{The $S$-parameter method}
\label{app:Smethod}

The normalised Stokes parameters can in principle be obtained from the
four HWP angles as $Q=[F(0\degr)-F(45\degr)]/2$ and
$U=[F(22.5\degr)-F(67.5\degr)]/2$, where
$F(\phi_i)=(f^o-f^e)/(f^o+f^e)$ at each angle \citep{Baglio2020}.
Since the polarisation degree $P=(Q^2+U^2)^{1/2}$ is not normally
distributed, and is biased upwards at low signal-to-noise
\citep{WardleKronberg1974}, we instead followed the $S$-parameter
method \citep[][and references therein]{Baglio2020}. For each HWP
angle we define
\begin{equation}
\label{eq:Sphi_i}
S(\phi_i) = \left(\frac{f^o(\phi_i)/f^e(\phi_i)}{f_u^o(\phi_i)/f_u^e(\phi_i)} - 1\right)
\Bigg/
\left(\frac{f^o(\phi_i)/f^e(\phi_i)}{f_u^o(\phi_i)/f_u^e(\phi_i)} + 1\right),
\end{equation}
where $f^o$ and $f^e$ are the ordinary- and extraordinary-beam fluxes
of the target, and $f_u^o$, $f_u^e$ the same quantities for a
reference field star, assumed to be intrinsically unpolarised. $S$ is
the component of the normalised Stokes vector along direction
$\phi_i$ (e.g.\ $S(0)=Q$), and is related to $P$ and to the
polarisation angle $\Phi$ by
\begin{equation}
\label{eq:Sphi}
S(\phi) = P\cos\!\big[2(\Phi-\phi)\big].
\end{equation}
Normalising by the reference stars removes, at once, the instrumental
polarisation and the interstellar polarisation common to the target
and the field. With $n$ reference stars, each HWP angle yields $n$
independent estimates of $S$, so that its distribution is sampled
directly rather than approximated as a Gaussian.

We fitted Eq.~(\ref{eq:Sphi}) to the full set of $(\phi_i, S)$ pairs
with a Bayesian approach, following \citet{Baglio2020}. A first
estimate of $P$ and $\Phi$ was obtained by maximising a Gaussian
likelihood with the Nelder--Mead algorithm \citep{GaoHan2012}. The
posterior probability density was then sampled with the
affine-invariant ensemble Markov-chain Monte Carlo sampler
\texttt{emcee} \citep{ForemanMackey2013,Hogg&Foreman2018}, with
walkers initialised in a small Gaussian ball around the best-fit
values. The first third of each chain was discarded as burn-in, and
we verified that a stationary distribution had been reached
\citep{Sharma2017}. We report the 0.16, 0.50, and 0.84 quantiles of
the marginalised posteriors as the credible interval and median of
$P$ and $\Phi$ \citep[cf.][]{Lucy2016}; for non-detections, we quote
the $3\sigma$ upper limit on $P$ derived from the same posterior.

\subsection{Instrumental polarisation and angle calibration}
\label{app:instpol}

The instrumental polarisation of FORS2 is monitored regularly by ESO
through observations of unpolarised standard stars, and has remained
stable at a low level ($0$--$0.3\%$) in all bands over the past decade
\citep{Baglio2020}. In any
case, the normalisation by the reference stars in
Eq.~(\ref{eq:Sphi_i}) removes any residual instrumental term along
with the interstellar one. The zero point of the polarisation angle was checked with the polarised
standard stars Hiltner~652 (2025 Aug~12/13) and BD$-$12\,5133
(Aug~22/23). In $R$, their measured angles ($178.3\pm0.3^\circ$ and $145.6\pm0.4^\circ$) are offset from the catalogue values ($179.39\pm0.10^\circ$ and $146.97\pm0.13^\circ$; \citealt{Fossati2007}) by $1.1\pm0.3^\circ$ and $1.4\pm0.4^\circ$, and in $I$ BD$-$12~5133 gives $142.9\pm0.6^\circ$ against $143.99\pm2.27^\circ$. The consistent sign and size of these offsets, on different nights, indicate a small systematic zero-point offset of $\simeq1.2^\circ$ (possibly including the slight wavelength dependence of the standards' angles between our filters and those of the catalogue) rather than underestimated uncertainties. This offset is smaller than the uncertainty on each individual epoch and would change the offset of $\overline{\Phi}_R$ from the \textit{IXPE} angle by $\sim1^\circ$, well within its \chg{$5.1^\circ$} uncertainty; we therefore did not apply it.

FORS2 also shows an off-axis instrumental polarisation, with a nearly axisymmetric pattern that increases from $\lesssim0.03\%$ on the optical axis to $\sim0.7\%$ at $3'$ from it in the $V$ band\footnote{\url{https://www.eso.org/sci/facilities/paranal/instruments/fors/inst/pola.html}}, and up to $\sim1.4\%$ at the edges of the field depending on the filter \citep{Patat2006,Gonzales2020}. Since the S-parameter normalisation assumes that the reference stars share the instrumental polarisation of the target, we selected reference stars within $\sim1.5'$ of it, where the effect is expected to be at most $\sim0.2\%$. Because the angle of this pattern follows the azimuth around the optical axis, its contribution partly cancels when averaging stars at different positions; moreover, the tight clustering of the reference stars in the $Q$--$U$ plane on every night (Appendix~\ref{app:ism}) shows that any residual off-axis term is small compared with our uncertainties.

\subsection{Reference stars and interstellar polarisation}
\label{app:ism}

The $S$-parameter method assumes that the reference stars are either
intrinsically unpolarised or polarised only by the same interstellar
medium as the target. We selected isolated, unsaturated field stars
detected in all images, six of which were usable on each
single-epoch night. Their Gaia DR3 \citep{GaiaMission,GaiaDR3} parallaxes
all lie within $\varpi\simeq0.22$--$0.45$\,mas, corresponding to
distances of $\sim\!2.2$--$4.5$\,kpc, which bracket that of
SAX~J1808.4$-$3658 ($3.5\pm0.1$\,kpc; \citealt{Galloway2006}). The
reference stars therefore sample an interstellar column comparable to
that towards the target.

To verify that the reference stars are not intrinsically polarised,
we computed their normalised Stokes parameters and those of the
target, with no correction applied, and plotted them in the $Q$--$U$
plane for each night and band. In every case the
reference stars cluster tightly around a common point, whose offset
from the origin measures the combined instrumental and interstellar
polarisation along the line of sight, while the target lies
separated from them. Table~\ref{tab:ism} lists this field
polarisation, from the weighted mean of the reference stars, for each
night and band. It is consistent between the three nights within
$\sim\!1\sigma$ in $R$ and within $\sim\!2.5\sigma$ in $I$.

The field polarisation is lower in $I$ than in $R$ on every night, as
expected for interstellar dust \citep{Serkowski1975}. The pooled
$R$-band value, $P_{\rm ISM}\simeq1.2\%$, lies below the maximum
interstellar polarisation expected for the extinction towards the
source, $P_{\rm max}\lesssim3A_V\simeq1.5\%$ for $A_V=0.51\pm0.04$
\citep{Serkowski1975,Baglio2020}. Both the level and the angle are
also in line with the 2019 campaign, in which the reference-star
polarisation was $\lesssim1\%$ at an angle of $\sim\!150\degr$ in all
bands, and a Serkowski fit gave $P_{\rm max}=1.18\pm0.08\%$ at
$\lambda_{\rm max}=0.44\pm0.08\,\mu$m \citep{Baglio2020}; our pooled
angles, $\chi_{\rm ISM}\simeq-27\degr\equiv153\degr$, are essentially
identical. The stability of the interstellar term across two outbursts
six years apart gives us further confidence that the reference stars
provide a robust baseline.

\begin{table*}
\caption{Field polarisation from the weighted mean of the reference
stars, per night and pooled.}
\label{tab:ism}
\centering
\begin{tabular}{lcccc}
\hline\hline
Night & $N_\star$ & $P_{\rm ISM}$ (\%) & $\chi_{\rm ISM}$ (\degr) & $(Q,U)$ (\%)\\
\hline
\multicolumn{5}{c}{$R$ band}\\
Aug 11/12 & 6 & $1.29\pm0.10$ & $-25.2\pm2.3$ & $(0.82,-1.00)$\\
Aug 20/21 & 6 & $1.16\pm0.07$ & $-28.6\pm1.7$ & $(0.63,-0.98)$\\
Sep 4/5   & 6 & $1.21\pm0.15$ & $-26.7\pm3.6$ & $(0.72,-0.97)$\\
Pooled    & 18 & $1.20\pm0.05$ & $-27.4\pm1.3$ & $(0.69,-0.98)$\\
\hline
\multicolumn{5}{c}{$I$ band}\\
Aug 11/12 & 6 & $1.00\pm0.08$ & $-24.8\pm2.3$ & $(0.65,-0.76)$\\
Aug 20/21 & 6 & $0.79\pm0.07$ & $-26.8\pm2.5$ & $(0.47,-0.63)$\\
Sep 4/5   & 6 & $1.13\pm0.12$ & $-32.9\pm2.8$ & $(0.46,-1.03)$\\
Pooled    & 18 & $0.87\pm0.05$ & $-26.6\pm1.5$ & $(0.52,-0.70)$\\ % weighted mean of the per-night values: replace with the pooled-fit output
\hline
\end{tabular}
\end{table*}

\subsection{Pooled reference-star baseline}
\label{app:pool}

Since the interstellar and instrumental polarisation of the field is
not expected to vary from night to night, the correction becomes more
accurate if the reference stars of all nights are combined, rather
than using only the six available on a single night. The consistency of the per-night values in Table~\ref{tab:ism} justifies
this choice, which we adopted for the three single-epoch nights, pooling their reference-star measurements (last row of each band in
Table~\ref{tab:ism}) before fitting Eqs.~(\ref{eq:Sphi_i}) and
(\ref{eq:Sphi}). The correction is computed separately for each band.
Pooling reduced the credible intervals on $\Phi$ by roughly a factor
of two relative to the single-night fits, and left $P$ consistent
within $1\sigma$ at each epoch. 
The 11-epoch timing sequence, being a single continuous observation, retains its own single-night reference-star baseline.

\subsection{Total fluxes from the polarimetric images}
\label{app:polflux}

Total fluxes of the target were obtained from the polarimetric images
themselves: for the target and the
reference stars, the $o$ and $e$ fluxes were summed at each HWP angle,
and the sums were averaged over the four angles. These instrumental
fluxes were then calibrated against the same field stars in the
contemporaneous LCO images, whose magnitudes are provided by the
\textsc{XB-NEWS} calibration (Appendix~\ref{app:lco}); $i'$-band
magnitudes were converted into the Cousins $I$ band with the
transformations of \citet{Jordi2006}. 
Fluxes were de-reddened with $A_V=0.51\pm0.04$, derived from the X-ray column density \citep{DiSalvo2019} and the $N_{\rm H}$--$A_V$ relation of \citet{Foight16}, and the extinction law of \citet{Gordon2023} (as implemented in \texttt{dust\_extinction}, \citealt{Gordon2024}; see also \citealt{Decleir_2022}), evaluated at the effective wavelengths of the Cousins $R$ and $I$ bands (0.641 and 0.798\,$\mu$m), and converted to flux densities with the corresponding Vega zero points of \citet{Bessell1998}.
Since the target flux is measured in the FORS2 images themselves and the
field stars are not variable, this calibration is not affected by the
LCO and VLT data not being simultaneous. However, the $R$ and $I$
exposures of each night are consecutive rather than simultaneous, and the
optical flux of SAX~J1808.4$-$3658 varies on timescales of minutes (by up
to $\sim\!0.15$\,mag in $\sim\!6$\,min in 2019; \citealt{Baglio2020}), so
the $R$ and $I$ fluxes of the same night can differ by up to this amount
for reasons unrelated to the spectral shape. This affects the
polarised-flux SED of Appendix~\ref{app:polSED}.

\subsection{Polarised-flux SED}
\label{app:polSED}

\begin{figure}
    \centering
    \includegraphics[width=1\linewidth]{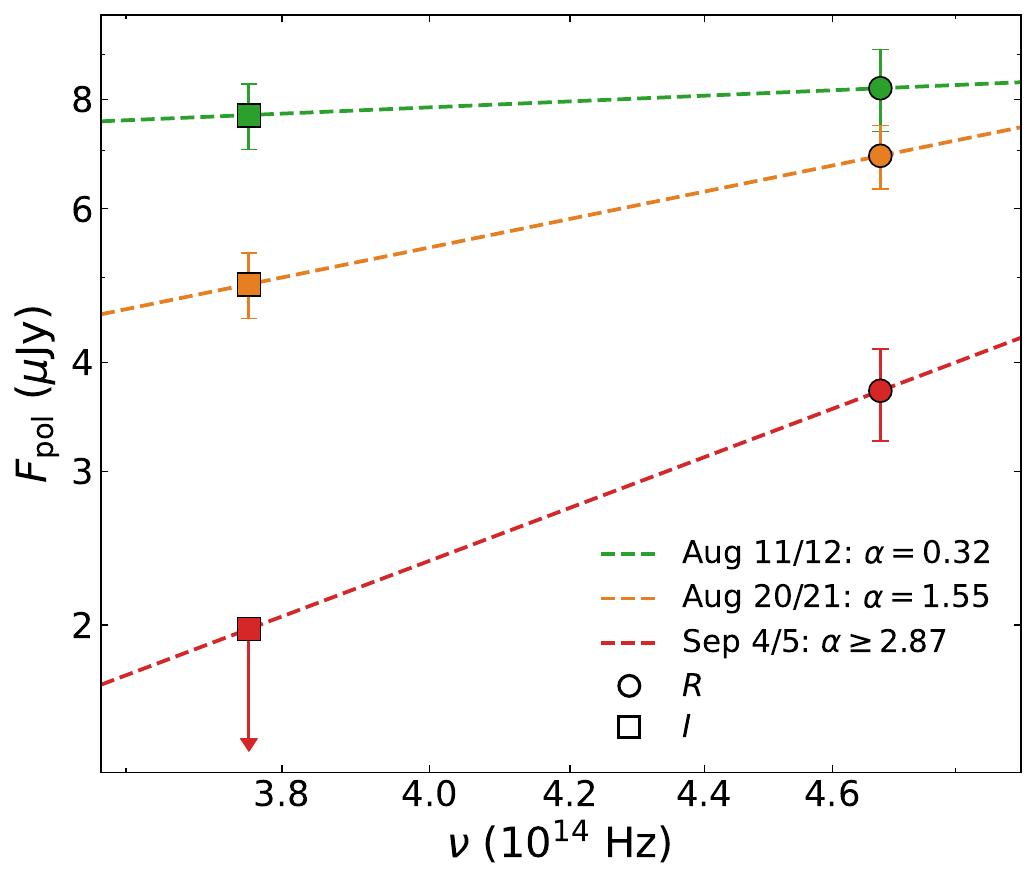}
    \caption{Polarised-flux SED of the three multi-band epochs, colour-coded as in Fig.~\ref{fig:phase} ($R$: circles; $I$: squares). Dashed lines: power laws ($F_{\rm pol}\propto\nu^\alpha$) through the $R$ and $I$ points of each epoch, with $\alpha$ given in the legend. For Sep~4/5 the $I$-band point is a $3\sigma$ upper limit and $\alpha$ a $3\sigma$ lower limit. Fluxes are de-reddened and converted at the effective wavelengths of the Cousins $R$ and $I$ bands (Appendix~\ref{app:polflux}).}
    \label{fig:polSED}
\end{figure}

If the polarisation arose from disc-atmosphere scattering, which varies
smoothly on the disc's dynamical and viscous timescales, the
polarised-flux spectral index would in principle be a physically
interpretable quantity. Combining the polarisation degrees of
Table~\ref{tab:results} with the de-reddened total fluxes
(Appendix~\ref{app:polflux}), we obtain the polarised-flux SED of the
three multi-band epochs (Fig.~\ref{fig:polSED}). It is not red at any of them and becomes progressively bluer as the outburst evolves: $\alpha=0.32^{+0.62}_{-0.64}$ at Aug~11/12, $\alpha=1.54^{+0.55}_{-0.54}$ at Aug~20/21, and $\alpha\geq2.88^{+0.51}_{-0.60}$ ($3\sigma$ lower limit) at Sep~4/5.

At Sep~4/5 we combine the $R$-band detection ($P=1.54^{+0.18}_{-0.19}\%$) and the conservative $I$-band $3\sigma$ upper limit ($P<0.98\%$; the $I$-band detection itself is only $\sim\!2\sigma$) with the de-reddened total fluxes ($F_R=0.241\pm0.001$\,mJy, $F_I=0.202\pm0.002$\,mJy), obtaining $F_{\rm pol,R}=3.72\,\mu$Jy and $F_{\rm pol,I}<1.98\,\mu$Jy, and hence $\alpha\geq2.88^{+0.51}_{-0.60}$ (conservatively, $\alpha\geq2.28$). This exceeds both the total-flux continuum slope ($\alpha_{\rm cont}=0.81$) and the disc blackbody slope at the CMD-implied temperature ($\alpha_{\rm BB}\simeq1.0$), and even the Rayleigh--Jeans limit ($\alpha=2$), the steepest slope a thermal spectrum, or electron-scattered thermal light, can reach.

\citet{Baglio2020} found a comparably extreme index of the opposite sign at their epoch~6 (2019, outburst decline): $P_R<0.16\%$ and $P_I=1.60\pm0.12\%$, implying a steeply negative index by the same method.
That the same source yields both extremely red and extremely blue
polarised-flux indices strongly suggests that polarised-flux SEDs built
from non-simultaneous exposures are unreliable in a system whose
polarisation flickers this fast.

%%%%%%%%%%%%%%%%%%%%%%%%%%%%%%%%%%%%%%%%%%%%%%%%%%%%%%%%%%%%%%
\section{LCO monitoring photometry}
\label{app:lco}

\subsection{Observations}
\label{app:lcoobs}

SAX~J1808.4$-$3658 has been monitored with the LCO network since 2008,
as part of a monitoring programme of $\sim$50 Galactic X-ray binaries
coordinated by the Faulkes Telescope Project \citep{Lewis2008}. During
the 2025 outburst, images were obtained with the 2\,m Faulkes
Telescope South and the 1\,m network nodes in the Bessell $V$
and $R$ and SDSS $i'$ filters. The field is very
crowded, and when the source is faint (near quiescence) it can be
blended with nearby stars, depending on the seeing
\citep{Baglio2020}; during the outburst, however, the target is
disentangled in the vast majority of images.

\begin{figure*}
\centering
\includegraphics[width=\hsize]{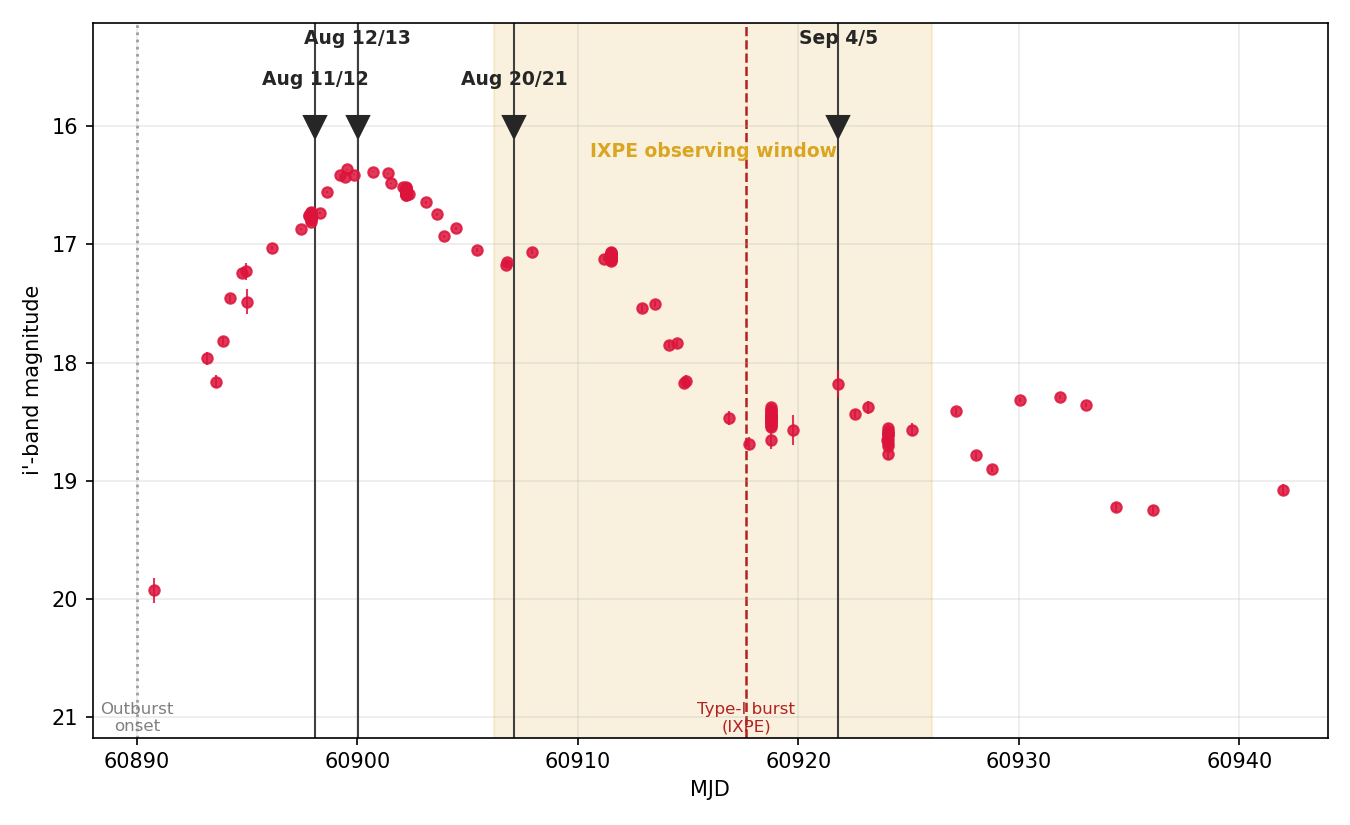}
\caption{LCO $i'$-band light curve of the 2025 outburst. The beige band
marks the \textit{IXPE} window, the grey dotted line the outburst onset, the red dashed line the Type-I burst, and black solid lines the four VLT/FORS2 polarimetric epochs. Where not visible, error bars are smaller than the symbols.}
\label{fig:lc}
\end{figure*}

The resulting $i'$-band light curve of the 2025 outburst, with the
VLT/FORS2 epochs marked, is shown in Fig.~\ref{fig:lc}.

\subsection{The XB-NEWS pipeline}
\label{app:xbnews}

The data were reduced with the X-ray Binary New Early Warning System
(\textsc{XB-NEWS}; \citealt{Russell2019,Saikia2026}) pipeline. The pipeline automatically downloads
the reduced images and their calibration data, and performs several
quality-control steps. For each image, sources are detected with
\textsc{SExtractor} \citep{Bertin1996}, and an astrometric solution
is computed with \textsc{astrometry.net} \citep{Lang2010} against
\textit{Gaia} DR2 positions. If the target is not detected within
$1''$ of its catalogued position, detection is repeated at a lower
threshold; if still undetected, forced photometry is performed at the
known target position.

Photometry is carried out with both multi-aperture photometry
\citep[MAP;][]{Stetson90} and fixed-aperture photometry, with
aperture radii scaled to an appropriate multiple of the PSF FWHM.
Light curves are then built by cross-matching detections across
images with the DBSCAN clustering algorithm \citep{Ester96}. The
photometry is calibrated against an extended version of the ATLAS
Reference Catalog 2 \citep[ATLAS-REFCAT2;][]{Tonry2018}, which
incorporates Pan-STARRS DR1 \citep{Chambers2016} and APASS DR10
\citep{Henden2018}. A photometric model including spatially variable
zero points, PSF-based terms, and source-specific mean magnitudes is
fitted to the light curves of all matched and unmatched sources
\citep{Bramich2012}, iteratively and with outlier
down-weighting to mitigate the effect of intrinsic variability. No
colour terms are included, which introduces only a small ($<1$--$2\%$)
systematic error in the absolute calibration. For the SDSS $i'$
filter, calibration uses the Pan-STARRS1 $i_{\rm P1}$ standard
magnitudes (AB system); $V$ and $R$ are calibrated indirectly via
standard magnitudes predicted for ATLAS-REFCAT2 stars from their
Pan-STARRS1 $g_{\rm P1}$ and $r_{\rm P1}$ magnitudes, using the
transformations of \citet{Tonry2012}. Forced-photometry measurements
with uncertainty $>0.25$\,mag are discarded as unreliable.

\subsection{Colour-magnitude diagram and disc model}
\label{app:cmd}

For the colour-magnitude diagram (Fig.~\ref{fig:cmd}), we used nightly-averaged $V$ and $i'$ magnitudes, pairing the two bands when taken within 1\,d. Because the optical flux varies on minute timescales (Appendix~\ref{app:polflux}), individual pairs taken hours apart can scatter significantly around the nightly trend; averaging reduces, but does not remove, this effect.

The disc track is a single-temperature, constant-area blackbody
heating and cooling \citep{MaitraBailyn2008,Russell2011}, applied to
SAX~J1808.4$-$3658 as in \citet{Baglio2020}. For a given extinction
($A_V=0.51\pm0.04$), the observed $V-i'$ colour is converted into an
intrinsic spectral index, and hence a blackbody temperature. At low temperatures the optical bands sample the curved peak of the
blackbody (and, at the coolest, its Wien side), while at higher
temperatures they move onto the Rayleigh--Jeans tail, where the
colour saturates, producing the characteristic shape of the track. The normalisation depends on
several poorly constrained quantities -- the disc size (set by the
orbital period and component masses), the disc filling factor, the
inclination (and possible warping), and the distance -- so, rather
than fitting it, \citet{Baglio2020} scaled it to reproduce the initial
optical brightening of the 2019 outburst, when the disc heated from
$\sim\!7000$ to $\sim\!10\,000$\,K \citep{Goodwin2020}. We adopt the
same model and normalisation here, so that the 2019 and 2025 outbursts
are directly comparable.

Departures of the data from the track towards redder colours trace an
additional red component, which in this source is attributed to
optically thin synchrotron emission from the jet
\citep{Baglio2020}. Following \citet{Baglio2020}, we quantify it as
$(V-i')_{\rm jet}$, the difference between the observed $V-i'$ colour
and that of the disc model, which is the
quantity plotted in Fig.~\ref{fig:V-i}.

\section{A jet launched from a warped inner disc}
\label{app:warp}
\begin{figure}
\centering
\includegraphics[width=\hsize]{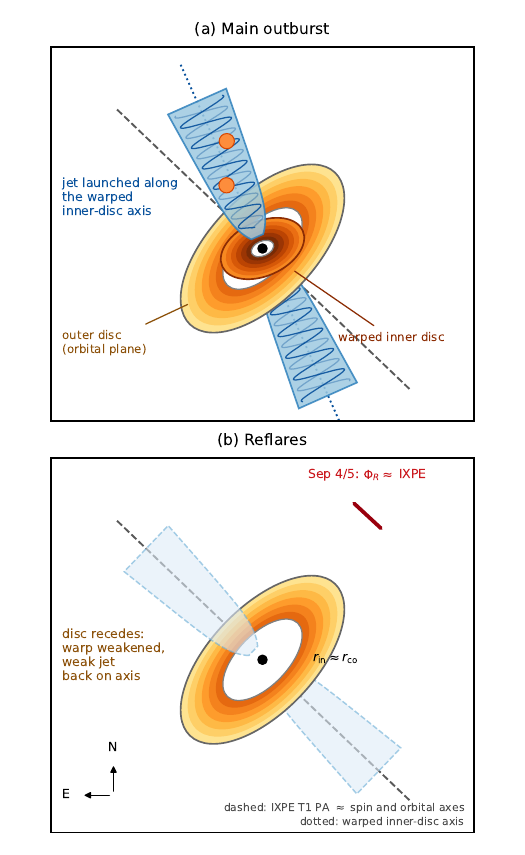}
\caption{Sketch of the warped-inner-disc alternative (not to scale;
North up, East left). (a) Main outburst: the jet is launched along the
axis of an inner disc warped out of the orbital plane (dotted), offset
from the \textit{IXPE} angle (dashed). (b) Reflares: as the inner disc
recedes towards $r_{\rm co}$, the magnetic warping weakens, and the
weak jet returns towards the axis.}
\label{fig:warp}
\end{figure}

An alternative origin for the offset of the net jet angle from the
spin axis (Sect.~\ref{sec:discussion}) is a jet launched along the axis
of an inner disc warped out of the orbital plane (Fig.~\ref{fig:warp}).
Unlike a misalignment of the whole disc, which the long accretion
history is expected to erase \citep{Lamb2009}, such a warp arises at the
inner disc edge, where it most directly affects the jet base, and is
continuously driven by the magnetic field: for an oblique dipole, the
magnetic torque exerted on the inner disc can drive warping and
precession, and a warping instability can push the disc plane away from
the stellar equatorial plane towards large tilts
\citep{Lai1999,PfeifferLai2004}, including for weakly magnetised
($\sim\!10^8$\,G) neutron stars such as SAX~J1808.4$-$3658
\citep{Shirakawa2002}. Radiation-driven warping \citep{Pringle1996}
instead acts on the outer disc, and is not expected in a system with
such a short orbital period \citep{Ogilvie2001}. Since the optical
synchrotron emission arises close to the jet base, any tilt would have
to be set at launch rather than acquired downstream. In this picture,
the X-ray polarisation, produced in the accretion columns, would still
trace the projected spin axis, since inside the magnetospheric radius
the flow is channelled by the neutron-star dipole regardless of the
orientation of the disc it leaves, while a disc-powered jet launched
from the warped region would follow the tilted inner-disc axis; a
spin-powered jet \citep{Parfrey2016} would instead remain
aligned with the spin. The offset of the main-outburst optical angle
would then favour disc-powered launching, consistent with the
evidence that slower jets, in both black holes and neutron stars,
change direction or precess and are therefore launched from the
accretion flow \citep{FenderMotta2025}. Whether a warp of
$\sim\!10$--$25\degr$ can be sustained depends on the magnetic
obliquity, which drives the warping torque; the small obliquity
inferred for this source ($\sim6\degr$; P26) makes such a large warp
unlikely. Moreover, a magnetically driven warp exists only
while the disc reaches the magnetosphere, and must re-form at each
outburst: the persistence of the same mean angle in 2019 and 2025 would
then require it to re-form with the same orientation, whereas it
follows naturally from a fixed geometric offset such as that of a wide
jet base (Sect.~\ref{sec:discussion}).

The scenario makes two testable predictions. First, during the
reflares the inner disc recedes towards $r_{\rm co}$, where the magnetic
warping torque, which falls steeply with radius \citep{Lai1999}, is
weaker. A jet launched from the inner disc would then return towards
the orbital axis, which coincides with the spin axis if the two are
aligned; this would make the alignment of $\Phi_R$ with the
\textit{IXPE} angle at Sep~4/5 a natural outcome rather than a
coincidence, if the jet still dominates the polarisation at that
epoch. Second, a magnetically driven warp precesses on the
short timescales of the inner disc \citep{Shirakawa2002}, unlike
tidally or radiation-driven warps of the outer disc, which precess over
years to decades. Precession faster than our 30--70\,s exposures would
average the jet angle onto the spin axis, so the persistent offset we
observe requires a quasi-static warp. Polarimetry on timescales of
seconds would reveal a periodic swing of the angle if the warp
precesses, distinct from the stochastic flickering expected from
compact emitting regions in a helical field.

\section{Emission at the disc--magnetosphere boundary}
\label{app:magneto}
SAX~J1808.4$-$3658 is one of the few accreting millisecond pulsars
with detected optical and ultraviolet pulsations: during the 2019
outburst, SiFAP2 detected optical pulsations with an rms amplitude of
$(0.55\pm0.06)\%$ and HST/STIS ultraviolet pulsations at
$(2.6\pm0.7)\%$ \citep{Ambrosino2021}, and UV and optical pulsations
were detected again in 2022 (\citealt{Ballocco2026_UV, Ballocco2026arXiv}). Their luminosity cannot be explained by reprocessing
or by cyclotron emission in the accretion columns \citep{Ambrosino2021}.
In the transitional millisecond pulsar PSR~J1023+0038, the optical
pulsations and polarisation are attributed to synchrotron emission
where the pulsar wind meets the inner disc, and the polarised and
pulsed fluxes coincide \citep{Papitto2019,Veledina2019,Baglio2025}.
In SAX~J1808.4$-$3658, however, the conditions differ: the X-ray
pulsations show that the accretion flow enters the magnetosphere,
whose radius ($r_{\rm m}\sim20$\,km at
$L_{\rm X}\sim10^{36}$\,erg\,s$^{-1}$ for $B\sim10^8$\,G) lies well
inside the light cylinder ($R_{\rm LC}=c/2\pi\nu\simeq120$\,km), and the X-ray luminosity, even when measured close in time to our Sep~4/5 reflare epoch, exceeds the spin-down power of the pulsar
\citep{Illiano2023,Papitto2026}.
The optical pulsations of SAX~J1808.4$-$3658 have been ascribed to
synchrotron radiation from particles accelerated where the disc
interacts with the magnetosphere, a process that can coexist with the
accretion-powered X-ray pulsations \citep{Ballocco2026arXiv}.

If this component also carries an ordered polarisation
(Fig.~\ref{Fig:combined}b), it could contribute to the polarisation
measured at Sep~4/5. In PSR~J1023+0038 the constant polarisation angle
over the spin cycle favours a predominantly poloidal field in the
emitting region \citep{Baglio2025}. For such a field, anchored to the
rotating neutron star, the angle would be fixed relative to the
projected spin axis, either parallel or perpendicular to it depending
on the field orientation; the alignment with \textit{IXPE} would
require the former. During the reflares the inner disc recedes
($r_{\rm m}\propto\dot M^{-2/7}$), in the spirit of the propeller and
trapped-disc pictures discussed in Sect.~\ref{sec:discussion}
\citep{Patruno2016}. A larger relative contribution of this component,
as the optical emission from the disc and jet fades, could then raise
$P$ and bring $\Phi$ onto the \textit{IXPE} angle, as observed.
Whether synchrotron emission from this region is optically thin at
optical wavelengths, as required for a significant polarisation,
remains to be shown.

The fluxes are at least compatible with this possibility. In 2019, the
optical pulsed flux during the main outburst was
$\simeq1.8\times10^{-14}$\,erg\,cm$^{-2}$\,s$^{-1}$ in the
320--900\,nm band \citep{Ambrosino2021}. This is comparable to the
$R$-band polarised flux we measure at Sep~4/5
($F_{\rm pol,R}=3.72\,\mu$Jy, i.e.
$\nu F_\nu\simeq1.7\times10^{-14}$\,erg\,cm$^{-2}$\,s$^{-1}$;
Appendix~\ref{app:polSED}). Because these measurements refer to
different outbursts and outburst stages, a common origin cannot be
established from them alone. A comparison with SiFAP2 pulsed fluxes obtained close in time
to our Sep 4/5 epoch is ongoing and will be presented in a forthcoming paper. Strictly simultaneous optical
polarimetry and fast photometry would provide a direct test. The
upcoming fast photopolarimeter eSiFAP, which can measure the
polarisation of the optical pulsations, would resolve the polarisation
at the spin period.

\end{appendix}
\end{document}